\documentclass{jfm}

\usepackage{graphicx}
\usepackage{natbib}

\usepackage{color}
\usepackage{amsmath}
\usepackage{subfigure}
\usepackage{calc}
\usepackage{ifthen}
\usepackage{tabularx}
\usepackage{contour}
\usepackage{varwidth}

\ifCUPmtlplainloaded \else
  \checkfont{eurm10}
  \iffontfound
    \IfFileExists{upmath.sty}
      {\typeout{^^JFound AMS Euler Roman fonts on the system,
                   using the 'upmath' package.^^J}       \usepackage{upmath}}
      {\typeout{^^JFound AMS Euler Roman fonts on the system, but you
                   dont seem to have the}       \typeout{'upmath' package installed. JFM.cls can take advantage
                 of these fonts,^^Jif you use 'upmath' package.^^J}             }
  \else
      \fi
\fi

\ifCUPmtlplainloaded \else
  \checkfont{msam10}
  \iffontfound
    \IfFileExists{amssymb.sty}
      {\typeout{^^JFound AMS Symbol fonts on the system, using the
                'amssymb' package.^^J}       \usepackage{amssymb}         \let\leq=\leqslant
         
      }{}
  \fi
\fi

\ifCUPmtlplainloaded \else
  \IfFileExists{amsbsy.sty}
    {\typeout{^^JFound the 'amsbsy' package on the system, using it.^^J}     \usepackage{amsbsy}}
    {}
\fi

  \newcommand\Rey{\mbox{\textit{Re}}}

\newsavebox{\astrutbox}
\sbox{\astrutbox}{\rule[-5pt]{0pt}{20pt}}

\newcommand{\der}{\mathrm{d}}

\newlength\tikzheight
\newlength\tikzwidth
\newlength\tikzwidthhalf
\newlength\xl
\newlength\yl
\newlength\zl
\makeatletter{}

\title[Global effect of local skin friction drag reduction in turbulent boundary layer]{Global effect of local skin friction drag reduction in spatially developing turbulent boundary layer}

\author[A. Stroh, Y. Hasegawa, P. Schlatter and B. Frohnapfel]{A. Stroh$^1$  \thanks{Email address for correspondence: alexander.stroh / bettina.frohnapfel @kit.edu},\ns
Y. Hasegawa$^2$\break
P. Schlatter$^3$
and B. Frohnapfel$^1$
}

\affiliation{
$^1$Institute of Fluid Mechanics, Karlsruhe Institute of Technology, Karlsruhe, Germany\\[\affilskip]
$^2$Institute of Industrial Science, The University of Tokyo, Tokyo, Japan\\[\affilskip]
$^3$Linn\'{e} FLOW Centre, KTH Mechanics, Stockholm, Sweden
}

\pubyear{2015}
\volume{650}
\pagerange{119--126}
\date{?; revised ?; accepted ?. - To be entered by editorial office}

\begin{document}

\maketitle

\begin{abstract}
\makeatletter{}A numerical investigation of two locally applied drag reducing control schemes is carried out in the configuration of a spatially developing turbulent boundary layer (TBL). 
One control is designed to damp near-wall turbulence and the other induces constant mass flux in the wall-normal direction.
Both control schemes yield similar local drag reduction rates within the control region. 
However, the flow development downstream of the control significantly differs: persistent drag reduction is found for the uniform blowing case whereas drag increase is found for the turbulence damping case.
In order to account for this difference the formulation of a global drag reduction rate is suggested.
It represents the reduction of the streamwise force exerted by the fluid on a finite length plate.  
Furthermore, it is shown that the far downstream development of the TBL after the control region can be described by a single quantity, namely a streamwise shift of the uncontrolled boundary layer, i.e. a changed virtual origin.
Based on this result, a simple model is developed that allows relating for the local drag reduction rate to the global one without the need of conducting expensive simulations or measurements far downstream of the control region. 
\end{abstract}


\section{Introduction}
\makeatletter{}Drag reducing flow control is a topic of great interest due to its practical significance for engineering application in different high-speed transport systems such as airplanes, marine vessels or pipelines.
A broad variety of control methods aiming at the reduction of skin friction drag in turbulent wall bounded flows has been introduced in the past. 
Classical active control schemes such as opposition control \citep{Choi_1994}, suboptimal and optimal control techniques \citep{Choi_1993,Lee_1998,Bewley_2001}, various wall movements  \citep{Quadrio_2011} and direct damping of near-wall fluctuations \citep{Iwamoto_2005,Frohnapfel_2010} have been thoroughly investigated during the last 20 years. 
These control schemes are often investigated using direct numerical simulations (DNS) of fully developed turbulent channel flow (TCF) assuming periodic boundary conditions in stream- and spanwise directions with control activated on the entire area of both channel walls. 
Although the TCF configuration has been a longstanding proven tool for the evaluation of control effects on the flow field, the global effect of locally applied control in a developing boundary layer is of major interest for the potential application of flow control in practice. 
This point is highlighted by  \cite{Spalart_2011} who present an analytic estimation of the drag reducing effect due to laminarization near the leading edge of a TBL. They draw attention to the fact that local and global control effects need to be distinguished when a TBL is locally altered.

Recently, several studies, in which different control strategies are analyzed in TBL, have been presented. 
In principle drag reduction in TBL can be achieved not only through the suppression of Reynolds shear stress, which is the typical strategy in TCF, but also by application of other control methods.
For example, the introduction of wall-normal mass flux is another drag reducing option \citep{Mickley1957}, which is thoroughly discussed based on DNS results in \cite{Kametani_2011} and \cite{Kametani_2015}. 
\cite{Stroh_2015} present a comparison of opposition control that aims at the suppression of the Reynolds shear stress in TCF and TBL and show that in spite of the very similar drag reduction rates that are achieved in both flows, the underlying drag reduction mechanisms are quite different.

Transient effects in TBL have sometimes been reported in literature. However, the main focus of these publications is not on the streamwise development of the TBL after the control and the corresponding computational domains are rather short so that the reported results are limited to a narrow region right after the control. 
\cite{Park_1999} conducted a DNS of TBL with local uniform blowing and suction and report a rapid decrease of the skin friction coefficient in the blowing region and a strong increase further downstream leading to negative local drag reduction rates. 
Similar results are reported by \cite{Kim_2002}.
\cite{Pamies_2007} present large eddy simulation data comparing locally applied uniform blowing, opposition control and blowing-only opposition control techniques. Downstream of the control they observe local drag increase for the blowing strategies and drag reduction for classical opposition control over a short region before the uncontrolled state is recovered. 
Oscillating wall control in TBL is discussed by \citet{Yudhistira_2011} and \citet{Lardeau_2013}. The former study analyzes the transient behavior at the beginning of the control region and the latter reports a reduction of the local skin friction drag downstream of the control area. 
Considering that a local flow state in spatially developing flow is generally determined as a result of the upstream events, it is possible that a local control affects the flow state  far downstream of the control region. This is an important aspect for potential practical applications, where the realization of a control technique on the entire wall might not be feasible, so that global drag reduction has to be achieved with a limited control area size.

In the present work we focus on the downstream development of a TBL and the resultant wall friction after control has been applied locally. 
This is achieved through DNS of TBL with sufficiently large streamwise domain length after a controlled region.
In the control region, drag reduction is achieved by either applying uniform blowing or by damping near-wall turbulence. 
The latter is a simplified representation of all classical TCF control techniques that lead to a suppression of the Reynolds shear stress. 
Both control schemes are adjusted such that similar drag reduction rates are realized in the control region. 
However, the downstream development of friction drag is quite different depending on the type of control applied. 
We propose the definition of a global drag reduction rate for TBL that is based on the overall  drag acting on a plate of finite length and show that this quantity is strongly influenced by the drag properties downstream of the control region. 
While the drag reduction achieved in the control region is specific to the applied control strategy, the streamwise development of friction drag far downstream of the control region can generally be explained by the von K\'{a}rm\'{a}n integral momentum equation. 
Based on this knowledge we propose a simple model for the drag estimation far downstream of the control region. 
 
\section{Numerical Procedure}
\makeatletter{}\begin{table}
\begin{center}
\def~{\hphantom{0}}
\begin{tabular}{ccccccccc}
\footnotesize{grid size}  &  \footnotesize{domain dimensions}  & \multicolumn{3}{c}{\footnotesize{resolution}} & \\
{$N_x\times N_y \times N_z$} &  {$L_x\times L_y \times L_z$} & {$\Delta x^+$} & {$\Delta y^+$} & {$\Delta z^+$}& {$\frac{L_y}{\delta_{99,max}}$} & $x_0$ & $\Delta x_c$ \\
\hline
{$3072 \times 301 \times 256$} &  {$3703 \times 123 \times 148$} & {$17.8$} & {$0.06-10.5$} & {$9.4$} & $2.3$ & $131$ & $247$  \\
\end{tabular}
\end{center}
\caption{
Domain properties and control configuration parameters. 
A non-dimensionalization based on $\theta \left( x_t \right)$ and $U_\infty$ is utilized (outer scaling), while viscous units (inner scaling) are based on the mean friction velocity in the turbulent region of the uncontrolled TBL. 
Dealiasing using the $3/2$-rule is employed in the $x$ and $z$ directions.
\label{table:config}}
\end{table}

The investigation is performed using DNS of a TBL under zero pressure gradient condition. 
The Navier-Stokes equations are numerically integrated using the velocity-vorticity formulation by a spectral solver with Fourier decomposition in the horizontal directions and Chebyshev discretization in the wall-normal direction \citep{Chevalier_2007}.
For temporal advancement, the convection and viscous terms are discretized using a 3rd order Runge-Kutta and Crank-Nicolson methods, respectively.
A weak random volume forcing in wall-normal direction close to the inlet plane is utilized for the boundary layer tripping and the fringe region technique is employed near the domain outlet.
In the fringe region the flow is forced to the laminar Blasius boundary layer profile with $\mathrm{Re}_{\delta_0^*} = 450$ based on the inflow displacement thickness $\delta_0^*$.
The tripping is located $10 \delta_0^*$ downstream of the inlet plane and is prescribed by attenuation lengths of $4 \delta_0^*$ and $\delta_0^*$ in streamwise and wall-normal directions with the temporal cutoff scale of $4 \delta_0^* / U_\infty$ and spanwise cutoff scale of $1.7 \delta_0^*$.
Adaptive time stepping is enabled during the simulations resulting in the average viscous time step of $\Delta t^+=0.15$ in the uncontrolled simulation.
The code and numerical domain configuration correspond to the ones reported in \cite{Schlatter_2009} and \cite{Schlatter_2012}.
Table~\ref{table:config} summarizes the numerical properties of the present simulations.

\begin{figure}
\centerline{\subfigure[\label{fig:scheme1}numerical domain]{
{\includegraphics{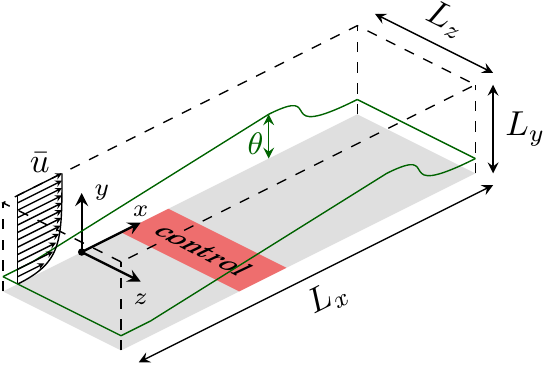}}
  }
  \hfill 
  \subfigure[\label{fig:scheme2}control placement]{
{\includegraphics{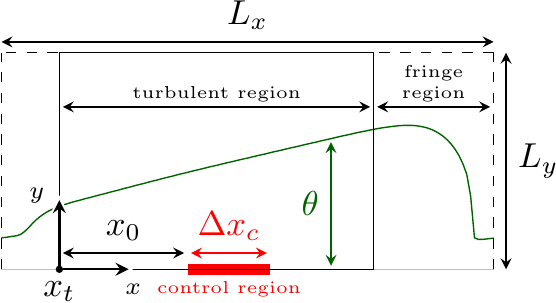}}
  }
  }
\caption{Schematic of the setup and applied control configuration.\label{fig:scheme}}
\end{figure}

Figure~\ref{fig:scheme1} demonstrates the general configuration of the numerical experiment with a local control region in the TBL and introduces a coordinate system with $x$, $y$ and $z$ representing streamwise, wall-normal and spanwise coordinate axes, respectively. 
The velocity components are denoted by $u$, $v$ and $w$, correspondingly.
The present configuration enables the development of a TBL up to $\Rey_\theta = 2500$ for the uncontrolled case with $\Rey_\theta = U_\infty \theta / \nu$.
Throughout the paper we consider only the turbulent region of the flow as shown in Figure~\ref{fig:scheme2}.
The momentum thickness at the beginning of the turbulent region, $\theta \left( x_t \right)$, and the free-stream velocity at the upper boundary, $U_\infty$, are used for the non-dimensionalization of all quantities. 
The beginning of the turbulent region is located $97$ length units inside the numerical domain, which corresponds to a local Reynolds number of $\Rey_\theta=365$.
In the following, the origin of the $x$-coordinate is placed at $x_t$.
This is different from the streamwise coordinate used in the definition of $\Rey_x = U_\infty x^* / \nu$, where $x^*$ marks the distance from the virtual origin of the uncontrolled TBL.
For the current simulation the position is estimated to be $289$ length units upstream of $x_t$~\citep{Bobke_2016}.
$\Rey_\theta$ of the uncontrolled case, $x$ normalized by $\theta \left( x_t \right)$ and $\Rey_x$ are used as alternative streamwise coordinates of the simulation domain.
Statistical integration is performed during a time period of at least $40 \delta_{99} / u_\tau$ at $\Rey_\theta=2500$ after the controlled flow reached an equilibrium state. 
This corresponds to approximately $32000$ viscous time units.

The skin friction drag is evaluated in terms of the dimensionless friction coefficient 
\begin{equation}
c_f \left(x \right) = \frac{\tau_w \left(x\right)}{\frac{1}{2} \rho U_\infty^2} ,
\end{equation}
which is a function of $x$ for a spatially developing flow. Based on this local skin friction drag  a local drag reduction rate is defined as
\begin{equation}
r \left( x \right) = 1 - \frac{\tau_w\left( x \right)}{\tau_{w,0}\left( x \right)}  = 1 - \frac{c_f\left( x \right)}{c_{f,0}\left( x \right)},
\label{eq:r}
\end{equation}
where the subscript of $0$ denotes values of the uncontrolled case. 

In order to assess the global turbulent drag reducing effect along a plate of finite length we integrate the local skin friction coefficient in streamwise direction from the beginning of the turbulent region $(x = 0)$ to a certain streamwise location $x$
\begin{equation}
\left[ c_f \right]_x = \frac{1}{ \frac{1}{2} \rho U_\infty^2} \int_{0}^x \tau_w \left( x\right) \mathrm{d}x =  \int_{0}^x c_f \left( x\right) \mathrm{d}x,
\label{eq:cfint}
\end{equation}
and define a global turbulent drag reduction rate based on this integral parameter (assuming the laminar and transitional development upstream of $x=0$ remains unchanged)
\begin{equation}
R\left(x\right) = 1 - \frac{ [c_f]_x }{[c_f]_{x,0}}.
\label{eq:rint}
\end{equation}
It should be noted that the global drag reduction rate $R\left(x\right)$ is not identical to the streamwise integral of the local drag reduction rate $r\left(x\right)$ since the skin friction coefficient of the uncontrolled flow changes in streamwise direction.

Two active control methods with different mechanisms leading to drag reduction are considered, namely, uniform blowing at the wall and body force damping of wall-normal fluctuations. 
Control is applied locally in streamwise direction, while the spanwise extent of the control region covers the entire domain width.
Both control types are placed at the same position, $x_0 = 131$, within the turbulent region with the same control region extension, $\Delta x_c = 247$, as shown in Figure~\ref{fig:scheme2}. 
This corresponds to a Reynolds number range of $\Rey_\theta=470-695$ or $\Rey_\tau=186-262$ based on $u_\tau$ of the uncontrolled case.
The location is defined by the control input profile:
\begin{equation}
f(x) = \begin{cases} 1, & \mbox{for } x_0 \leq x \leq  x_0+\Delta x_c \\ 0,  & \mbox{otherwise.} \end{cases}
\end{equation}
Additionally, the control amplitude is smoothly increased and decreased within a spatial extent of $14$ length units at the edges inside the control region using a hyperbolic tangent function in order to prevent the Gibbs phenomena.

For  uniform blowing the control input is constant in time and defined as
\begin{equation}
v_w \left( x \right) = \alpha \cdot f\left( x \right),
\end{equation}
where $v_w$ is the wall-normal velocity at the wall and $\alpha$ represents the blowing intensity.
Body force damping is based on the control algorithms proposed by \cite{Iwamoto_2005} and \cite{Frohnapfel_2010}. 
The control law aims at the suppression of wall-normal fluctuations in the near-wall region and utilizes the wall normal velocity component as sensor information. 
The control input is given as body force in $y$-direction:
\begin{equation}
b_y(x,y,z,t)=- \frac{f \left(x \right) \cdot g \left( y \right)}{\Phi} \cdot v(x,y,z,t) ,
\label{eq:bf}
\end{equation}
where  $\Phi$ is a time constant of the forcing that determines the relative strength of the applied body force.
The body force is applied up to $y=2.5$ such that
\begin{equation}
g(y) = \begin{cases} 1, & \mbox{for } 0 \leq y \leq 2.5 \\ 0,  & \mbox{otherwise.} \end{cases}
\end{equation} 
The control amplitude is smoothly decreased in wall-normal direction within $1.5 < y < 2.5$.
Using viscous units based on the local wall shear stress of the uncontrolled TBL the body force is activated in the region up to $y^+\approx45$.

Both control schemes are adjusted to yield a similar global drag reduction rate at the end of control region. 
In order to achieve this, the blowing intensity is set to $\alpha = 0.005 U_\infty$ and the forcing time constant to $\Phi = 5/3$.
The blowing intensity is similar to the intensities investigated in \cite{Kametani_2011}, \cite{Kametani_2015} ($0.001-0.010 U_\infty$) or \cite{Pamies_2007} ($0.0025 U_\infty$) and is significantly smaller than the intensities introduced in the earlier numerical studies by \cite{Park_1999} ($0.0185-0.0925 U_\infty$) or \cite{Kim_2002} ($0.0124-0.0463 U_\infty$).
For the considered simulations the present blowing intensity translates into the average blowing velocity $v_w^+=0.10$ based on the local $u_\tau$ of the uncontrolled case or $v_w^+=0.14$ when the local $u_\tau$ of the controlled case is considered.
The chosen control parameters result in $R(x_0+\Delta x_c)=33\%$ and $31\%$ for body force damping and uniform blowing, respectively.

\section{Results}
\makeatletter{}\begin{figure}
\begin{center}
\includegraphics[width=\textwidth]{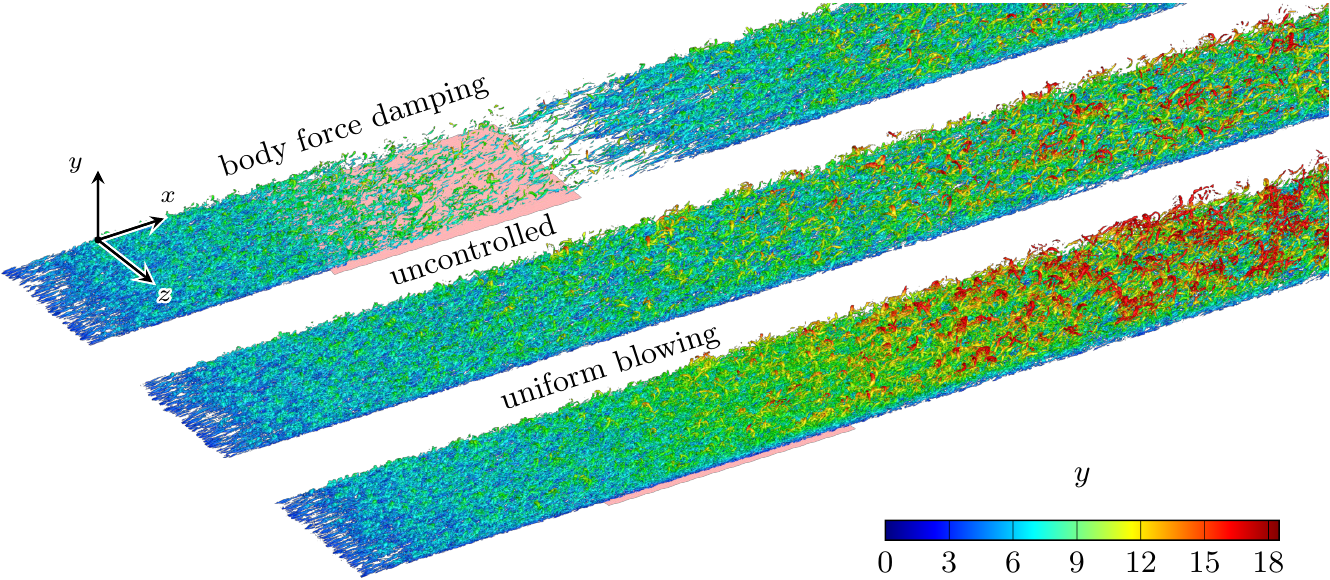}
\end{center}
\caption{Flow structure in uncontrolled and controlled cases represented by the isosurfaces of $\lambda_2$-criterion ($\lambda_2 = -0.005$) colored by the wall-normal coordinate. The red shaded area marks the location of the applied control. \label{fig:lambda2_comp}}
\end{figure}

Figure~\ref{fig:lambda2_comp} shows the influence of the applied control on the turbulent structures of the flow.
Due to cancellation of the wall-normal fluctuations in the near wall region a strongly pronounced attenuation of turbulence can be observed for body force damping. 
The effect is also visible over a certain extent downstream of the control region, where a re-transition of the flow occurs.
In contrast, the application of uniform blowing rather leads to visible thickening of the TBL due to additional wall-normal mass and momentum, which is accompanied by an enhancement of turbulent activity.

\subsection{Local and global turbulent drag reduction rates}
\begin{figure}
\begin{center}
\includegraphics{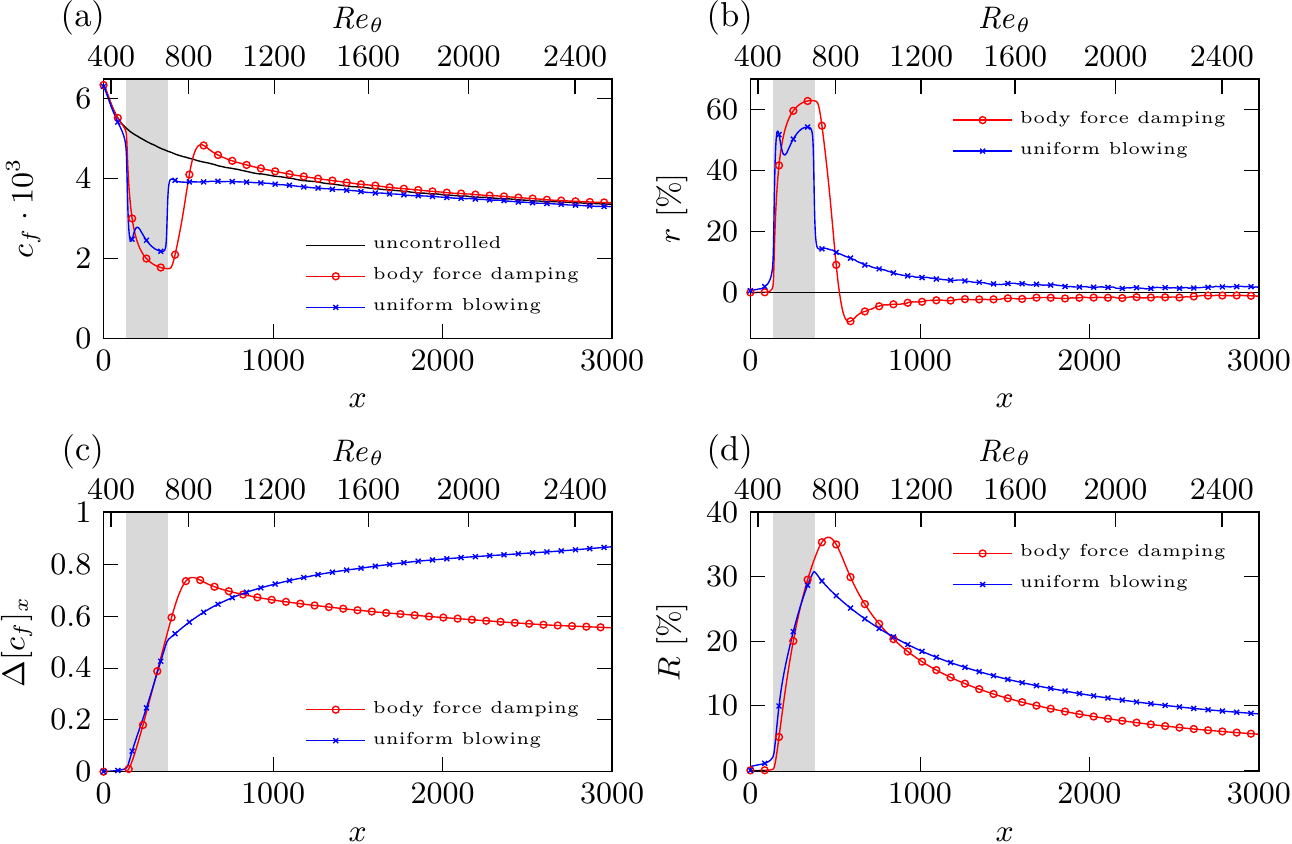}
\end{center}
\caption{Streamwise development of (a) local skin friction coefficient, (b) local drag reduction rate, (c) difference in integral skin friction coefficient and (d) global drag reduction rate. 
The shaded area marks the location of the control region. \label{fig:r_rint}}
\end{figure}

Figure~\ref{fig:r_rint}(a) shows the streamwise evolution of the skin friction coefficient for the considered control schemes.
The corresponding local drag reduction rate is depicted in Figure~\ref{fig:r_rint}(b).
Both schemes cause significant modification of the skin friction coefficient resulting in local drag reduction rates with maximum values of $63\%$ and $55\%$, respectively. 
For body force damping, drag reduction increases gradually inside the control region ($131 \leq x \leq 378$) and approaches the maximum of $r=63\%$ at the end of the control region. 
Downstream of the controlled region, $r$ first decays, reaches a negative peak, and then gradually converges to zero. 
This result is in agreement with \citet{Lardeau_2013}.
For uniform blowing, $r$ starts to increase upstream of the control region already. 
It peaks at $r=53\%$, decreases inside the control region and rises back to $55\%$ at the end of the control region. 
This local decrease of $r$ inside the control region is also found in \citet{Park_1999} and \citet{Pamies_2007}. 
A positive $r$ remains downstream of the control region, which gradually decays towards zero. 
This downstream behavior is in a good agreement with the results reported by \citet{Kametani_2015}, but it differs from the data presented in \citet{Park_1999} and \citet{Kim_2002} where negative $r$ is reported downstream of the blowing section.
This apparent contradiction can be solved through a simple parameter variation, which reveals that the downstream behavior directly after the control section strongly depends on the control configuration. 
Results of such a parameter variation are shown in Appendix~\ref{appA}.
It is found that a shorter control region with the same blowing amplitude causes a local negative $r$ directly after the control.
This is in agreement with the observations of \citet{Park_1999} and \citet{Kim_2002}, where only a short region after the control is employed due to a limited streamwise domain size.
The larger domain size in the present DNS reveals that - also in these cases - $r$ assumes positive values far downstream of the control region. Persistent drag reduction far downstream of the control is therefore a characteristic property of uniform blowing.

The  evolution of $r$ downstream of the control region significantly influences the integral turbulent skin friction drag and the global turbulent drag reduction rate when a longer section of the flat plate after the control region is taken into consideration.
The corresponding results for $\Delta [c_f]_x = [c_{f,0}]_x - [c_{f}]_x$ and $R$ are shown in Figure~\ref{fig:r_rint}(c) and (d), respectively.
The control schemes were set up to yield a similar global behavior within the control region, which is reflected in the figures. Both reach  $\Delta [c_f]_x \approx 0.5$ and $R \approx 30\%$ at the end of the control. 
In the case of uniform blowing $\Delta [c_f]_x$ continuously increases downstream of the control while body force damping produces a peak of $\Delta [c_f]_x$ around $x=525$ before it decreases continuously. 
In terms of $R$ it can be seen that for the present case body force damping will outperform uniform blowing if a plate up to a total length of $x\approx 800$ is considered. 
For a longer plate, uniform blowing yields better global behavior. 
It should be emphasized that the global drag reduction rate up to $x = 3000$ achieved by uniform blowing is $56\%$ larger than that by body force damping, although they yield very similar drag reduction rate within the control region. This is solely caused by the different behavior downstream of the control region.

For drag reducing control in TCF the formulation of additional control indices related to net energy savings or energy efficiency of the applied control \citep{Kasagi_2009} is widely established. 
This evaluation is based on the fact that any active control requires a certain power input to run the control. 
This continuously inserted control power has to be smaller than the reduction in pumping power due to drag reduction in a TCF in order to enable net energy savings. 
For TBL the dimensionless power saving obtained by the global drag reduction is given by $\Delta \left[c_f \right]_{x}$ shown in Figure~\ref{fig:r_rint}(c) multiplied with the free-stream velocity $U_\infty$, which is unity. 
An estimation of the control power input for uniform blowing requires the consideration of the pressure difference between the plenum from which the injected mass is supplied and the flow. For an idealized case in which this pressure difference is neglected, the power that the control inserts into the fluid is very small \citep{Kametani_2011}.   
In case of body force damping, the local power input due to the applied body force is given by the product of body force $b_y$ (as given by equation~\eqref{eq:bf}) and wall normal velocity fluctuation. 
Since the body force is designed to oppose the instantaneous velocity fluctuation this product is always negative. 
Therefore, the control actually extracts  energy from the turbulent flow field. 
Considering the fact that the wall normal velocity fluctuations are almost entirely damped to zero in the control region, the extracted power approaches very small values. For both control schemes in the present study, the power input to or the extraction from the flow field is so small that net energy savings rates are expected to be very similar to the drag reduction rates. 
While this holds for the considered idealized systems (negligible pressure difference for continuous blowing and loss-free control actuators) additional energy losses will be present for any non-ideal case but are not quantifiable without further specification of the control system. 

\subsection{Boundary layer development downstream of the control region}

\begin{figure}
\begin{center}
\includegraphics{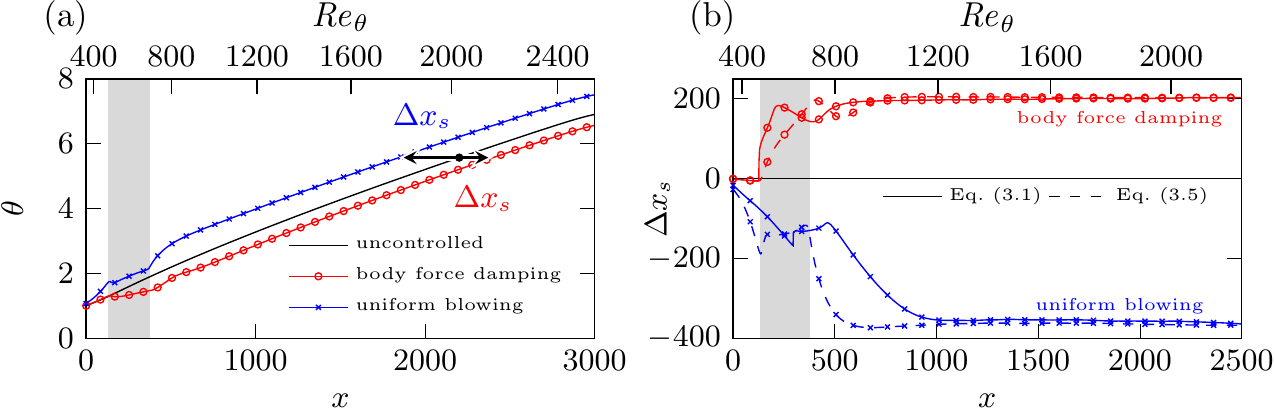}
\end{center}
\caption{Streamwise development of  (a) momentum thickness $ \theta$ and (b) estimation of the corresponding spatial shift $\Delta x_s$.
The shaded areas mark the location of the control region. \label{fig:mt_shift}}
\end{figure}

The persistent local drag change after the control region shown in Figure~\ref{fig:r_rint} suggests that the TBL recovers to the canonical state with a slightly different boundary layer thickness than it would have been at the same position in the uncontrolled case.
This is confirmed in Figure~\ref{fig:mt_shift}(a), where the spatial development of the momentum thickness for the two controlled cases are shown. 
Depending on the type of applied control, the momentum thickness at a certain streamwise position downstream of the control is either increased or decreased.
The suppression of turbulence with body force damping leads to reduced momentum loss in the controlled region and thus to less increase of the momentum thickness. In contrast, uniform blowing directly increases the boundary layer thickness. This difference in momentum thickness remains even after the flow returns to an uncontrolled equilibrium state. 
We therefore propose that the remaining effect far downstream of the control region can be described by a streamwise shift $\Delta x_s$ of the virtual origin of the TBL as indicated in Figure~\ref{fig:mt_shift}(a). The solid lines in figure Figure~\ref{fig:mt_shift}(b) show the evaluation of $\Delta x_s$ based on minimizing the difference between $\theta$ of the controlled case and $\theta_0$
\begin{equation}
\label{deltaxs}
\min \left(\theta_0 \left( x \right) - \theta \left( x +\Delta x_s \right) \right)^2 .
\end{equation}
It can be seen that the considered $\Delta x_s$ converges to a constant value  downstream of the control region. 
It assumes positive values for body force damping and negative ones for uniform blowing.  
This different shift is consistent with the long tails of negative and positive $r$ observed for body force damping and uniform blowing, respectively, far downstream of the controlled region (see Figure~\ref{fig:r_rint}).
Since the wall friction in the uncontrolled flow decreases monotonically with increasing streamwise distance from the leading edge, the positive/negative shift of the leading edge corresponds to an increase/decrease of local wall friction at a fixed streamwise position $x$. 
In this respect, uniform blowing is advantageous if long distances after the control are present because it yields not only a significant drag reduction in the controlled region, but also a greater momentum thickness (and thereby lower wall friction)  downstream of the control region due to a negative $\Delta x_s$.
This estimation is based on the assumption that the flow far downstream of the control region returns to a canonical state.

\begin{figure}
\hspace*{2.75pt} \includegraphics{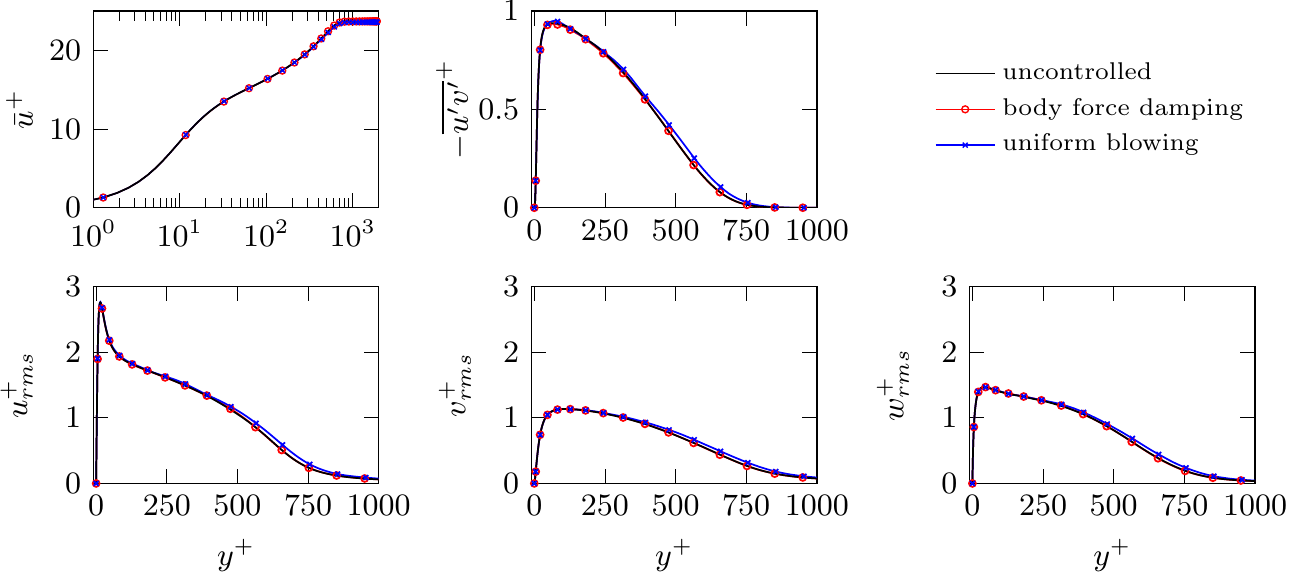}
\caption{
Flow statistics for uncontrolled and controlled TBL at the same momentum thickness ($Re_\theta = 1980$) downstream of the control region. The actual position corresponds to $x = 2125$, $x = 2125+199 = 2324$ and $x = 2125 - 356  = 1769$ for the uncontrolled flow, body force damping and uniform blowing, respectively.
\label{fig:stat}}
\end{figure}

The assumption that the flow returns to the same generic boundary layer state is justified in the following:
Figure~\ref{fig:stat} shows statistical properties of the uncontrolled flow and the two controlled flows shifted by the respective $\Delta x_s$ resulting in the same momentum thickness.
It can be seen that very good agreement is achieved in case of body force damping while small deviations exist for uniform blowing in the outer layer. 
These findings are in agreement with the results of \citet{Schlatter_2012} and indicate that in case of uniform blowing $\Delta x_s$ is not sufficient to produce a fully canonical state. 
In the present paper we choose to use the definition of $\Delta x_s$ as given by Eq.~\eqref{deltaxs} since we address properties of the mean flow only.

\subsection{Relation between local control and the resultant streamwise shift}
In order to investigate the control effect on the spatial development of the momentum thickness in more detail, we revisit the von K\'{a}rm\'{a}n integral momentum equation \citep{vonKarman_1946, Goldschmied_1951} describing the relationship between $\theta$ and the wall friction, based on the boundary-layer equations,
\begin{equation}
\frac{\der \theta}{\der x} = \frac{c_f}{2} + \frac{v_{w}}{U_\infty} + \frac{1}{\rho U_\infty^2} \int_0^\infty \frac{\partial p}{\partial x} \der y + \frac{1}{U_\infty^2} \int_0^\infty \frac{\partial \overline{u'u'}}{\partial x}  \der y,
\label{eq:karman}
\end{equation}
where $p$ is the static pressure and $\overline{u'u'}$ represents the variance around the local mean value of $u$.
Integration of this equation in streamwise direction (from the beginning of the turbulent region at $x=0$ to a point $x$ downstream of the controlled region)
leads to the following expression for $\theta$:
\begin{equation}
\theta \left( x \right)  \approx   \frac{1}{2} \left[ {c}_f \right]_x + \int_{0}^x \frac{v_w \left( x \right)}{U_\infty} \der x + 1.
\label{eq:karmanint}
\end{equation}

The last two terms of Eq.~(\ref{eq:karman}) are neglected because their streamwise integration results in a very small contribution to $\theta$ only. It should be noted that positive and negative local pressure gradients exist at the beginning and the end of the control region, respectively. Therefore, Eq.~(\ref{eq:karmanint}) can only be applied for estimations of $\theta$ far downstream of the controlled region. In the present case, the estimation based on Eq.~(\ref{eq:karmanint}) leads to deviations of less than $5\%$ if applied for $x>500$ in case of body force damping and $x>700$ for uniform blowing (while the end of the controlled region is located at $x=378$). These deviations decrease if longer integration areas are considered and are on the order of $1\%$ towards the end of the simulation domain. 

The remaining terms in Eq.~(\ref{eq:karmanint}) describe the far downstream evolution of $\theta$. 
The second term on the right-hand-side is proportional to the net-mass-flux from the wall up to a streamwise distance $x$, and is present only for (uniform) blowing. In the case of body force damping 
this term is always null. 
In this case, drag reduction - i.e. a lower $\left[ {c}_f \right]_x$ - always leads to a slower increase of the momentum thickness. For uniform blowing, however,  the momentum thickness
can be greater than the uncontrolled value despite the fact that a reduction of the first term, i.e., wall friction, is achieved because the second term is always positive. 

Although Eq.~(\ref{eq:karmanint}) can be used to estimate the change of the momentum thickness caused by an arbitrary local control, it is  not very convenient to describe the downstream development of the flow field since $\Delta \theta$ is a function of $x$. 
In this respect, it is more useful to convert the change in momentum thickness to the streamwise shift $\Delta x_s$ introduced in the previous section. 
An estimation of the streamwise shift from the change in momentum thickness can be performed using the empirical relationship between $\Rey_x$ and $\Rey_\theta$ proposed by \cite{Nagib_2007}:
\begin{equation}
\Rey_x = \frac{\Rey_\theta}{\kappa^2} \left( \left( \ln \Rey_\theta + \kappa B -1 \right)^2 +1 \right),
\end{equation}
with $\kappa = 0.41$ and $B = 5.2$ \citep{Pope_2000}, which provide a good agreement with the present simulation results.
The resulting relation between $\Delta x_s$ and $\theta$ downstream of the control region reads
\begin{equation}
\Delta x_s = \frac{\theta_0}{\kappa^2} \left( \ln \Rey_{\theta,0} + \kappa B - 1 \right)^2 - \frac{\theta}{\kappa^2} \left( \ln \Rey_{\theta} + \kappa B - 1 \right)^2 - \frac{\Delta \theta}{\kappa^2},
\label{eq:dxsemp}
\end{equation}
where $\Delta \theta = \theta_0-\theta$.

\begin{figure}
\begin{center}
\includegraphics{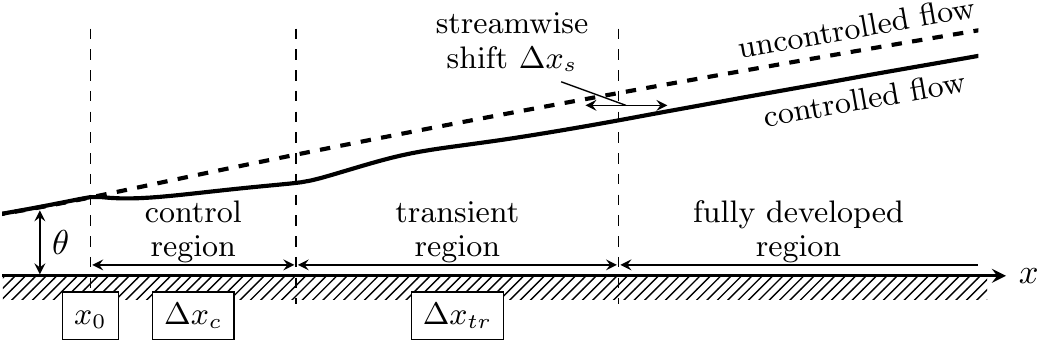}
\end{center}
\caption{
 Schematic of turbulent boundary layer subject to local control. \label{fig:regions}}
\end{figure}

In Figure~\ref{fig:mt_shift}(b), the estimate of $\Delta x_s$ based on Eq.~\eqref{eq:dxsemp} is also plotted. 
It can be seen that it agrees well with the DNS data if regions far enough downstream are considered. 
A schematic of TBL subject to local control is shown in  Figure~\ref{fig:regions}.
We name the region where an agreement with shifted uncontrolled solution is reached fully developed and refer to the region directly downstream of the control region as transient region. 
For the present cases we find a streamwise shift in the fully developed region of  $\Delta x_s = 204$ for body force damping and  $ \Delta x_s= -365$ for uniform blowing, respectively. 
The corresponding length of the transient region is given by $ \Delta x_{tr}= 500$ and $ \Delta x_{tr}= 750$.
The transient regions end at $\Rey_\tau=370$ and $540$ (based on local $u_\tau$) for body force damping and uniform blowing, respectively. 

\subsection{Influence of the control placement}

\begin{figure}
\begin{center}
\includegraphics{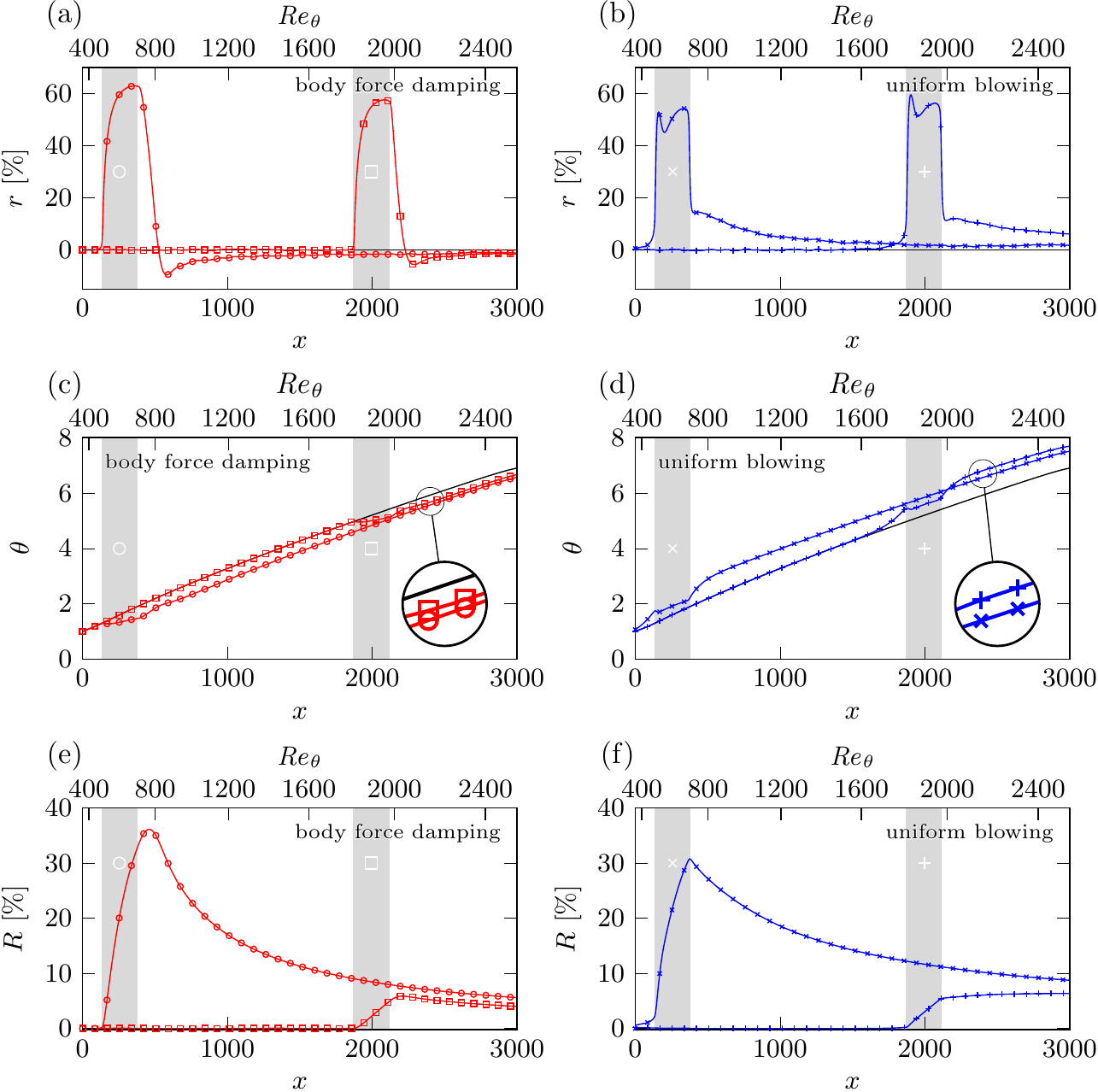}
\end{center}
\caption{Streamwise development of the local drag reduction rate, the momentum thickness and the integral drag reduction rate with different control placement for (a,c,e) body force damping and (b,d,f) uniform blowing. The shaded area marks the location of the control region. \label{fig:r_int_placement}}
\end{figure}

Local and global drag reduction rates are also influenced by the exact placement of the control. 
A parametric study with varying location of the control region reveals that the application of drag reducing control closer to the leading edge of the TBL is generally more beneficial in terms of integral drag reduction despite the fact that the local drag reduction rates show opposite trends for the two control configurations.

As an example, Figure~\ref{fig:r_int_placement} presents a comparison for different control placements. 
The beginning of the control region is placed at $x_0 = 131$ and $x_0=1870$, while all other control parameters are kept the same in physical units; i.e. $\Delta x_c=247$, turbulence damping is applied up to $y=2.5$ and the injected mass flow rate is fixed.
For the cases where the control is placed farther downstream, the corresponding Reynolds number range of the controlled region is $\Rey_\theta=1810-1976$ or $\Rey_\tau=624-674$ based on $u_\tau$ of the uncontrolled case.
It should be noted that the same physical control parameters correspond to different values in local viscous units, i.e. when body force damping is placed farther downstream the control is activated in the region up to $y^+ \approx 38$ and for uniform blowing the average blowing velocity $v_w^+=0.118$ based on $u_\tau$ of the uncontrolled case is present.
Figures~\ref{fig:r_int_placement}(a,b) show the local drag reduction rate for these cases.

For body force damping the local control performance is slightly reduced in the case when the control is placed farther downstream ($57\%$ maximum drag reduction instead of $63\%$). A similar effect is known for TCF where $r$ decreases for increasing friction Reynolds number when control is applied in the same near wall region in terms of viscous units \citep{Iwamoto_2002}. For TBL $u_\tau$ slightly decreases with distance from the leading edge while the boundary layer thickness significantly increases. Therefore, a placement further downstream corresponds to a placement at higher friction Reynolds number. The reduction of $r$ with increasing $Re_\tau$ is not surprising in the sense that the ratio of the wall-normal height of the control region to the boundary layer thickness is smaller; 
i.e. $20\%$ at $x_0 = 131$ and $5\%$ at  $x_0=1870$. Although the general decrease of $r$ with increasing Reynolds number for wall based skin friction drag reducing techniques is well known for TCF, the exact relationship is subject of present investigations \citep{Gatti_2015}. 

In contrast, for uniform blowing the local control performance is slightly increased for farther downstream positioning of the control ($59\%$ maximum drag reduction instead of $55\%$).  
This difference can be understood if one considers the injected mass flow rate in viscous units. 
The viscous length scales decrease with $x$ such that in viscous units the injected mass flow rate is larger at the further downstream position. 
Therefore, the resulting change in skin friction drag is larger. 

Since the local change of skin friction drag directly affects the change of the momentum thickness far downstream, as shown in Figures~\ref{fig:r_int_placement}(c,d), the control placement farther downstream results in a smaller positive $\Delta \theta$ for body force damping and a larger negative $\Delta \theta$ for uniform blowing.
With respect to the global drag reduction rate Eq.~\eqref{eq:karmanint} can be used to establish  a relationship between $R$ and the introduced $\Delta \theta$:
\begin{equation}
R \left( x \right) = \left( \Delta \theta \left(x \right) + \int_{0}^x \frac{v_w \left( x \right)}{U_\infty} \der x  \right) \cdot \frac{1}{\theta_0 \left(x \right) -1}.
\label{eq:rinttheta}
\end{equation}
The momentum thickness at the end of the considered plate for uncontrolled flow conditions, $\theta_0(x)$, is a fixed value. For body force damping it holds that $v_w=0$ such that $R$ is solely determined by $\Delta \theta$. The reduction of a positive $\Delta \theta$ as discussed above therefore corresponds to lower values for $R$ for later control placement which is in agreement with the results shown in Figure~\ref{fig:r_int_placement}(e). In the case of constant blowing $v_w$ is constant such that differences in $R$ can also only arise from changes in $\Delta \theta$. The relation therefore suggests that the reported larger negative $\Delta \theta$ will also result in smaller $R$ values for later placement, independent of the total length of the plate that is considered. The results shown in Figure~\ref{fig:r_int_placement}(f) confirm that lower $R$ values are indeed found for the later placement. Actually, the reduction of $R$ is more pronounced for uniform blowing than for body force damping which is in agreement with the fact that larger changes in $\theta$ are introduced. 

From Figures~\ref{fig:r_int_placement}(e,f) it is evident that the streamwise development of $R$ directly after the control section can include a positive streamwise gradient ($R' > 0$) before $R$ decays continuously in the far downstream region: $R' > 0$ is found for body force damping (regardless of the control placement) and for the further downstream placement of uniform blowing. The streamwise gradient of $R$ can be described as
\begin{equation}
R' \left( x \right) = \left. \frac{\der R}{\der x} \right|_x  = \frac{c_{f,0} \left( x \right)}{ \left[c_f \right]^2_{x,0}} \cdot \left( r \left(x \right) \cdot \left[c_f \right]_{x,0} - \Delta \left[c_f \right]_{x} \right),
\label{eq:rstrmain}
\end{equation}
which reveals that the sign of $R'$ depends on the balance of the two terms in the brackets. In case of body force damping $r$ remains at large positive values over some distance after the control region yielding a dominance of the first term and thus $R'>0$. For uniform blowing, the first term dominates only when the control region is placed further downstream. This dominance disappears eventually and $R'<0$ is established for  $x > 2930$
which suggests that further downstream application of blowing will not outperform its upstream application in terms of $R$ for the chosen control parameters. However, we note that the monotonic decrease of $R$  far downstream of the control region as given by equation~\eqref{eq:rstrmain} does not provide a firm proof that the upstream control placement is the optimal one.
The detailed derivation of Eq.~(\ref{eq:rstrmain}) and further discussion can be found in Appendix~\ref{appB}.

\subsection{Estimation of global drag reduction rates achieved by local control}

\begin{figure}
\begin{center}
{\includegraphics{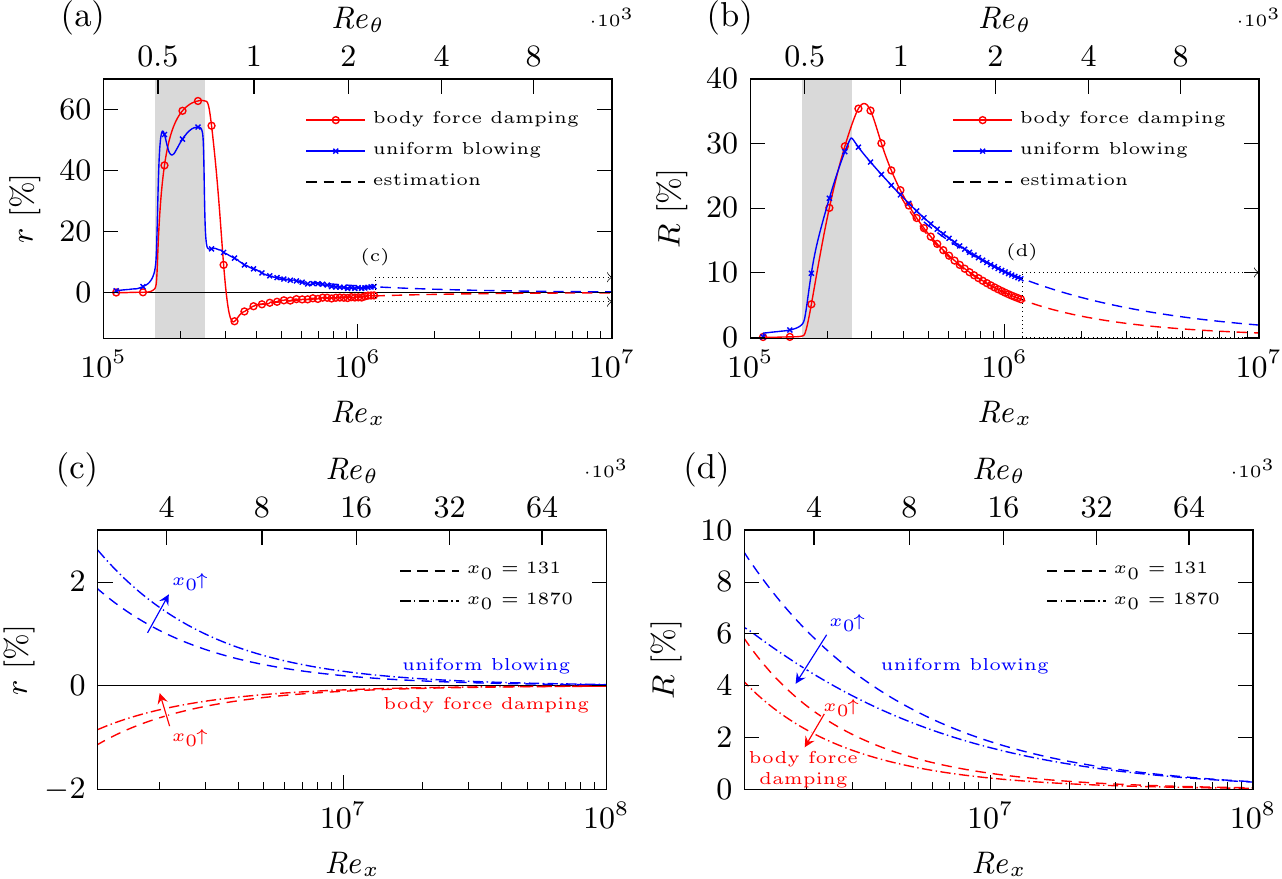}}
\end{center}
\caption{Comparison of the simulation data with the estimation of the downstream (a) local and (b) integral drag reduction rate based on the spatial shift and dependence of the estimated development on the control placement (c,d).
The shaded area marks the location of the control region at $x_0=131$. 
\label{fig:model}}
\end{figure} 
 
The spatial development of skin friction drag within the control and transient regions depends on the applied control scheme, and thus has to be evaluated via DNS or experiment. 
However, it has been shown that the far downstream effect of a control region in TBL can be characterized by a single quantity, namely  $\Delta x_s$, which represents a streamwise shift of the TBL virtual origin.
Therefore, the drag in this fully developed region can be estimated without relying on DNS or experiments in very large domains. 

Once the influence of drag reducing control on the TBL is reduced to  $\Delta x_s$  the development of $r$ and ${R}$ for the fully developed region can be estimated using empirical correlations between the skin friction coefficient and the Reynolds number, e.g. based on \cite{White_2006}
\begin{equation}
c_f = 0.4177 \left( \log \left( 0.06 \Rey_x \right) \right)^{-2}.
\label{eq:white}
\end{equation}
The streamwise coordinate $x'$ that enters the Reynolds number is the coordinate of the uncontrolled flow simply corrected by the shift such that $x'=x^*+ \Delta x_s^*$, where $\Delta x_s^*$ is the dimensional representation of $\Delta x_s$ and $x^*$ corresponds to the distance from the virtual origin of the uncontrolled solution.
Corresponding results are depicted in Figure~\ref{fig:model}. 
Figure~\ref{fig:model}(a) and (b) show an excellent agreement between the model adopted for the fully developed region of the TBL and the DNS data. 
The estimation can be used to further extrapolate the results and shows that local and global drag reduction rates will eventually vanish if long uncontrolled sections after the control region are considered.
Figure~\ref{fig:model}(c) and (d) present the prediction downstream of the simulation domain extended up to $Re_x = 10^8$ for the control regions placed at $x_0=131$ and $1870$.
Utilizing equation~\eqref{eq:dxsemp} with $\Delta \theta (x=3000)$ the streamwise shift introduced by the control region at $x_0=1870$ is estimated to be $\Delta x_s = 147$ and $-515$ for body force damping and uniform blowing, respectively.
Figure~\ref{fig:model}(c) shows a less pronounced permanent local drag increase for the body force damping and a higher local drag reduction for the uniform blowing when control is applied farther downstream.
However, for both control schemes higher integral drag reduction is observed when control is placed closer to the leading edge of the TBL as discussed in the previous section.


\section{Conclusions}
\makeatletter{}The global effect of locally applied skin friction drag reduction in TBL is addressed in this paper. 
We define a global drag reduction rate as an integral quantity that includes controlled and uncontrolled regions of the flow over a plate of finite length. 
In a TBL DNS two fundamentally different drag reducing control schemes are locally applied and the global drag reducing effect is evaluated with a particular focus on the downstream development after the control region. 
Although the control parameters are designed so that both control methods yield similar drag reduction rate inside the control region, the downstream development of the wall friction is found to be fundamentally different. The control scheme that attenuates turbulent activity leads to a thinner TBL downstream of the control region in comparison to the uncontrolled flow and thus yields local drag increase in this region. In contrast, the control scheme of local uniform blowing thickens the TBL downstream of local uniform blowing and therefore exhibits persistent drag reduction. 

The present results indicate that in a developing flow any local manipulation influences the downstream development via a streamwise shift of the uncontrolled solution.
While the local drag reduction within the control region depends on the type of applied control and the location at which it is applied, the control effect far downstream of the control region can be generally expressed by a streamwise shift of the virtual origin of the TBL.

Since the streamwise shift captures the global influence of the particular control technique on TBL, the quantity can also be understood as a control performance index for TBL.
The representation of the control effect through one single quantity allows predicting the drag behavior far downstream of the control without the need of DNS or experiments in this region.
For instance, such prediction enables an estimation of the distance over which a net drag reducing effect is observed when a large-eddy break-up device (LEBU) is utilized for TBL control as pointed out by \cite{Narasimha_1988}.
In general, the present results indicate that drag reducing control applied in the beginning of a TBL has a higher global efficiency in term of the suggested integral drag reduction rate. 
In addition, it is shown that any local drag reduction effect vanishes from a global perspective if long uncontrolled sections after the control region are present.

Similar downstream effects like the ones reported here are likely to appear if local wall roughness is considered. 
Skin friction drag will increase in the rough region but downstream of the rough region the increased boundary layer thickness will lead to lower local skin friction drag such that the global drag change strongly depends on the combination of rough and smooth regions along the plate.

Finally, it should be noted that for a plate of finite length the overall drag is not solely determined by skin friction. The wake of the plate, which depends on the boundary layer thickness at the trailing edge, will also contribute to the overall flow resistance. In addition, for aeronautical applications the applied control techniques might also influence the lift which is not included in the present consideration. 

\section*{Acknowledgements}
\makeatletter{}This work was performed on the computational resources bwUniCluster and ForHLR Phase I funded by the Ministry of Science, Research and the Arts Baden-W\"urttemberg and DFG (``Deutsche Forschungsgemeinschaft'') within the framework program bwHPC.
The authors greatly acknowledge the support by DFG through project FR2823 /2-1, and the Ministry of Education, Culture, Sports, Science and Technology (MEXT) Japan through Grand-in-Aid for Scientific Research (B) (No. 25289037).

\bibliographystyle{jfm}

\appendix
\makeatletter{}\section{}\label{appA}
\begin{figure}
\begin{center}
\includegraphics{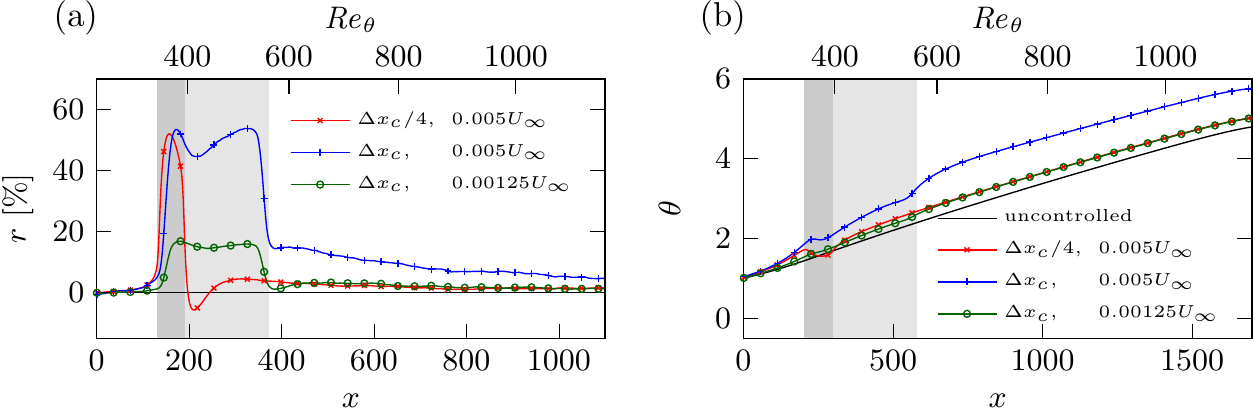}
\end{center}
\caption{Streamwise development of (a) local drag reduction rate and (b) momentum thickness for uniform blowing configurations with variation of control region length and blowing intensity. 
Shaded area marks the location of the control region. \label{fig:short}}
\end{figure}

A parametric study with variation of control length and blowing amplitude is carried out in a shorter simulation domain with TBL up to $\Rey_\theta = 1100$.
For consistency, the same $x_t$, $x_0$ and non-dimensionalization based on $\theta \left( x_t \right)$ of the main simulation are used in the parametric study.
As an example, three cases are considered: the base control configuration from the main simulation with $\Delta x_c=247$ and $v_w = 0.005 U_\infty$, a configuration with a shorter control length $\Delta x_c /4$ and $v_w = 0.005 U_\infty$ and a configuration with lower blowing amplitude $v_w = 0.00125 U_\infty$ over the same control length $\Delta x_c$. 
The latter two configurations exhibit the same bulk blowing.
In Figure~\ref{fig:short}(a) it can be seen that the shortest control length yields negative $r$ values right after the control region while the other two remain positive. 
This effect is presumably associated with the flow adjustment through a rapid enhancement of turbulent activity when a strong blowing is introduced over a short area, whereas a longer control area enables a gradual adaptation of the flow field to the introduced blowing so the region with negative $r$ vanishes.
Figure~\ref{fig:short} (b) shows the evolution of the momentum thickness for the considered control configurations.
It is evident that downstream of the control area the TBL is rendered thicker and hence yields lower skin friction coefficient with corresponding positive local drag reduction for all cases.

\section{}\label{appB}

Based on Eqs.~\eqref{eq:cfint} and \eqref{eq:rint} the streamwise gradient of $R$ at a certain location $x$ can be written as
\begin{equation}
R' \left( x \right) = \left. \frac{\der R}{\der x} \right|_x = \frac{\Delta \left[c_f \right]'_x}{\left[c_f \right]_{x,0}} + \Delta\left[c_f \right]_x \left( \frac{1}{\left[c_f \right]_{x,0}} \right)'.
\label{eq:rstr}
\end{equation}
Considering that
\begin{equation}
\Delta \left[c_f \right]'_x = \frac{\der \Delta \left[c_f \right]_x}{\der x} = c_{f,0} \left(x \right) - c_{f} \left(x \right),
\end{equation}
and
\begin{equation}
\left( \frac{1}{\left[c_f \right]_{x,0}} \right)' = -\frac{\left[c_f \right]'_{x,0}}{\left[c_f \right]^2_{x,0}} = -\frac{c_{f,0} \left( x \right)}{ \left[c_f \right]^2_{x,0}},
\end{equation}
the relationship~\eqref{eq:rstr} can be rewritten as
\begin{equation}
\begin{split}
R' \left( x \right) & = \frac{\left( c_{f,0} \left( x \right) - c_f \left( x \right) \right)}{\left[c_f \right]_{x,0}} - \frac{\Delta \left[c_f \right]_{x} \cdot c_f \left( x \right)}{\left[c_f \right]^2_{x,0}}
\\
& = \frac{\left( c_{f,0} \left( x \right) - c_f \left( x \right) \right) \cdot \left[c_f \right]_{x,0} - \Delta \left[c_f \right]_{x} \cdot c_{f,0} \left( x \right)}{\left[c_f \right]^2_{x,0}}
\\ 
& = \frac{\left( r \left( x \right) \cdot c_{f,0} \left( x \right) \right) \cdot \left[c_f \right]_{x,0} - \Delta \left[c_f \right]_{x} \cdot c_{f,0} \left( x \right)}{\left[c_f \right]^2_{x,0}}
\\
& = \frac{c_{f,0} \left( x \right)}{ \left[c_f \right]^2_{x,0}} \cdot \left( r \left(x \right) \cdot \left[c_f \right]_{x,0} - \Delta \left[c_f \right]_{x} \right),
\end{split}
\label{eq:derrstr}
\end{equation}
with $r \left( x \right)$ as defined in Eq.~\eqref{eq:r}

Since the prefactor of Eq.~\eqref{eq:derrstr} depends only on the properties of uncontrolled flow, the sign of $R'\left( x \right)$ is determined by the balance between the two terms in the parenthesis.
Assuming that local drag reduction is achieved in the control section, this balance reveals that the streamwise gradient of $R$ can be negative (second negative term is dominant) or positive (first term is dominant and positive).
The resultant $R' \left( x \right)$ depends on the control placement and the particular control type.
Due to the fact that $\left[c_f \right]_{x,0}$ monotonically increases with $x$ by definition and changes in $r \left( x \right)$ and $\Delta \left[c_f \right]_{x,0}$ are minor for different control placements, it is expected that the second term is dominant when the control is placed closer to the leading edge, while the first term is dominant when the control is placed further downstream.
In the case of body force damping, local drag increase is introduced downstream of the control section ($r \left( x \right)<0$), so $R' \left( x \right)$ is always negative (except the short section close to the control area where $r \left( x \right)$ remains positive).
On the other hand, uniform blowing introduces a permanent local drag reduction downstream of the control section ($r \left( x \right)>0$) and $R' \left( x \right)$ tends to be positive, especially when the control section is placed farther downstream. 

\end{document}